\documentclass[%
 reprint,
superscriptaddress,
 amsmath,amssymb,
aps,
]{revtex4-2}

\usepackage{graphicx}
\usepackage{dcolumn}
\usepackage{bm}
\usepackage{multirow}
\usepackage{makecell}
 \usepackage{float}

\usepackage{units}
\usepackage{color}
\usepackage{url}

\usepackage[colorlinks]{hyperref}
\hypersetup{%
	plainpages=true,
	breaklinks=true,
	hypertexnames=false,
	pageanchor=true,
	colorlinks=true,
	linkcolor={blue},
	citecolor={red},
	urlcolor={blue},
	anchorcolor={black}
}

\usepackage{mleftright} 

\newcommand{\figref}[1]{\mbox{Fig.~\ref{#1}}}
\newcommand{\figpanel}[2]{Fig.~\hyperref[#1]{\ref*{#1}(#2)}}
\newcommand{\figpanels}[3]{Fig.~\hyperref[#1]{\ref*{#1}(#2)-(#3)}}
\newcommand{\figpanelNoPrefix}[2]{\hyperref[#1]{\ref*{#1}(#2)}}

\renewcommand{\eqref}[1]{\mbox{Eq.~(\ref{#1})}}
\newcommand{\tabref}[1]{\mbox{Table~\ref{#1}}}
\newcommand{\appref}[1]{\mbox{Appendix~\ref{#1}}}
\newcommand{\bra}[1]{\mleft\langle #1 \mright |}
\newcommand{\ket}[1]{\mleft|#1 \mright \rangle}

\begin{document}

\preprint{APS/123-QED}

\title{Robust preparation of Wigner-negative states \\ with optimized SNAP-displacement sequences}

\author{Marina Kudra}
\email{kudra@chalmers.se}
\affiliation{Department of Microtechnology and Nanoscience, Chalmers University of Technology, 412 96 Gothenburg, Sweden}

\author{Mikael Kervinen}
\affiliation{Department of Microtechnology and Nanoscience, Chalmers University of Technology, 412 96 Gothenburg, Sweden}

\author{Ingrid Strandberg}
\affiliation{Department of Microtechnology and Nanoscience, Chalmers University of Technology, 412 96 Gothenburg, Sweden}

\author{Shahnawaz Ahmed}
\affiliation{Department of Microtechnology and Nanoscience, Chalmers University of Technology, 412 96 Gothenburg, Sweden}

\author{Marco Scigliuzzo}
\affiliation{Department of Microtechnology and Nanoscience, Chalmers University of Technology, 412 96 Gothenburg, Sweden}

\author{Amr Osman}
\affiliation{Department of Microtechnology and Nanoscience, Chalmers University of Technology, 412 96 Gothenburg, Sweden}

\author{Daniel P\'erez Lozano}
\affiliation{Department of Microtechnology and Nanoscience, Chalmers University of Technology, 412 96 Gothenburg, Sweden}

\author{Mats O. Thol\'en}
\affiliation{Nanostructure Physics, KTH Royal Institute of Technology, 114 19 Stockholm, Sweden}
\author{Riccardo Borgani}
\affiliation{Nanostructure Physics, KTH Royal Institute of Technology, 114 19 Stockholm, Sweden}
\author{David B. Haviland}
\affiliation{Nanostructure Physics, KTH Royal Institute of Technology, 114 19 Stockholm, Sweden}

\author{Giulia Ferrini}
\affiliation{Department of Microtechnology and Nanoscience, Chalmers University of Technology, 412 96 Gothenburg, Sweden}

\author{Jonas Bylander}
\affiliation{Department of Microtechnology and Nanoscience, Chalmers University of Technology, 412 96 Gothenburg, Sweden}

\author{Anton Frisk Kockum}
\affiliation{Department of Microtechnology and Nanoscience, Chalmers University of Technology, 412 96 Gothenburg, Sweden}

\author{Fernando Quijandr\'{\i}a}
\thanks{Present address: Quantum Machines Unit, Okinawa Institute of Science and Technology Graduate University, Onna-son, Okinawa 904-0495, Japan.}
\affiliation{Department of Microtechnology and Nanoscience, Chalmers University of Technology, 412 96 Gothenburg, Sweden}

\author{Per Delsing}
\email{per.delsing@chalmers.se}
\affiliation{Department of Microtechnology and Nanoscience, Chalmers University of Technology, 412 96 Gothenburg, Sweden}

\author{Simone Gasparinetti}
\email{simoneg@chalmers.se}
\affiliation{Department of Microtechnology and Nanoscience, Chalmers University of Technology, 412 96 Gothenburg, Sweden}

\date{\today}

\begin{abstract}

Hosting non-classical states of light in three-dimensional microwave cavities has emerged as a promising paradigm for continuous-variable quantum information processing. Here we experimentally demonstrate high-fidelity generation of a range of Wigner-negative states useful for quantum computation, such as Schr\"{o}dinger-cat states, binomial states, Gottesman-Kitaev-Preskill (GKP) states, as well as cubic phase states. The latter states have been long sought after in quantum optics and were never achieved experimentally before. We use a sequence of interleaved selective number-dependent arbitrary phase (SNAP) gates and displacements. We optimize the state preparation in two steps. First we use a gradient-descent algorithm to optimize the parameters of the SNAP and displacement gates. Then we optimize the envelope of the pulses implementing the SNAP gates. Our results show that this way of creating highly non-classical states in a harmonic oscillator is robust to fluctuations of the system parameters such as the qubit frequency and the dispersive shift.

\end{abstract}


\maketitle

\section{\label{sec:Introduction}Introduction}

A promising approach to realize a quantum computer is to encode the quantum information into bosonic modes~\cite{ma2021quantum,joshi2021quantum}. This approach has several advantages compared to a quantum computer based on two-level systems~\cite{divincenzo1995quantum}. First, the information is redundantly encoded in the infinite-dimensional Hilbert space of a harmonic oscillator. Choosing the encoding wisely, paired with a single dominant error channel, single-photon loss, makes quantum error correction possible in a hardware-efficient way~\cite{gertler2021protecting,ofek2016extending,hu2019quantum}. Moreover, non-Clifford gates can be implemented more efficiently on this platform~\cite{gottesman2001encoding,joshi2021quantum,bourassa2021blueprint}.

The basic building blocks of the bosonic quantum computer are nonclassical states generated in a harmonic oscillator. The nonclassicality of the states is conveniently characterised by the negativity of their Wigner function, which is a necessary resource for quantum computational advantage~\cite{Mari2012,veitch2012negative}. One way to generate these states is to resonantly exchange excitations between a frequency-tunable ancillary qubit and the oscillator~\cite{Law1996,hofheinz2008generation,hofheinz2009synthesizing}. This approach works well for states containing up to a few photons, thanks to fast swaps enabled by resonant interaction, but does not scale well beyond that, since one needs to climb the Fock-state ladder in a sequential manner to reach higher photon numbers. 

Another approach is to use optimal control pulses in a system comprising a fixed-frequency qubit and an oscillator in the strong dispersive regime~\cite{heeres2017implementing}. This approach works well, yielding high-fidelity states, because it addresses all the relevant transitions at once. However, it relies on very accurate estimation of the parameters of the model Hamiltonian, and fine-tuning of the drive parameters. In addition, it works like a ``black box'' in which the system dynamics at intermediate times cannot be used to gain insight into the mechanisms that limit the fidelity of the state preparation.
By contrast, a gate based approach using displacement and selective number-dependent arbitrary phase (SNAP) gates~\cite{heeres2015cavity,krastanov2015universal} offers a more transparent alternative. Not only are these two gates enough for universal control of the harmonic oscillator~\cite{krastanov2015universal}, but they can also, in theory, be applied in an efficient way to greatly reduce the number of required gates~\cite{fosel2020efficient}. A potential downside is that the standard implementation of the SNAP gates limits the achievable fidelity of the target states, because the duration of the pulses must be significantly longer than the inverse of the dispersive interaction between the qubit and the oscillator~\cite{heeres2015cavity}, which makes these gates slow and as a result more prone to decoherence effects.

In this work, we experimentally demonstrate high-fidelity generation of a variety of Wigner-negative states in a microwave cavity by applying up to three blocks of interleaved displacement and SNAP gates. We show that states used in error-correction protocols, such as the Fock, binomial~\cite{michael2016new,hu2019quantum}, and cat states~\cite{vlastakis2013deterministically,ofek2016extending}, can be prepared with fidelities between 0.93 and 0.99. We also show that complex quantum states enabling universal continuous-variable quantum computing when supplied to Gaussian circuits, such as the Gottesman-Kitaev-Preskill (GKP) state~\cite{gottesman2001encoding, baragiola2019, campagne2020quantum,fluhmann2019encoding} and the cubic phase state~\cite{gottesman2001encoding, Lloyd1999, hillmann2020,yanagimoto2020engineering}, can be generated. In particular, the cubic phase state has to our knowledge not been implemented before~\cite{yukawa2013emulating, miyata2016implementation}. In contrast to previous work \cite{heeres2015cavity, heeres2017implementing}, we optimize the pulse sequences with a two-step approach. First we optimize the parameters that define the applied SNAP and displacement gates using a variation of an algorithm proposed by F\"osel \textit{et al.}~\cite{fosel2020efficient}. Then we numerically optimize the envelope of individual SNAP pulses to decrease the pulse length by about a factor of 8, which results in a fivefold decrease in infidelity. Moreover, we show that optimizing the SNAP pulses makes these gates more robust to fluctuations or erroneous estimation of system parameters such dispersive shift and the qubit frequency, compared to SNAP gates based on pulses with a comb-like structure in frequency domain~\cite{heeres2015cavity}.

\begin{figure}[t]
\includegraphics[width=90mm]{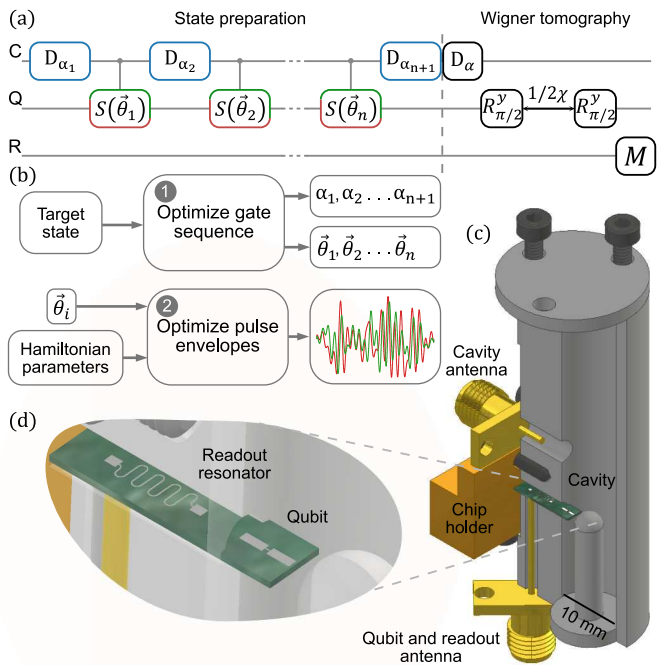}
\caption{\label{fig:setup} (a) Pulse sequence used to prepare and characterize the states. C, Q and R stand for cavity, qubit and readout resonator respectively. $D_{\alpha_i}$ are the displacement and $S(\vec{\theta}_i)$ are the SNAP gates. $R_{\pi/2}^y$ are cavity-state-independent $\pi/2$ pulses. They are separated in time by $1/2\chi$, where $\chi$ is the dispersive shift between the qubit and the cavity. $M$ is the measurement pulse. (b) The two steps of optimization that we perform to find the desired displacement and SNAP gates. (c) Drawing of the stub geometry cavity coupled to the transmon qubit. (d) Close-up of the chip containing the stripline readout resonator and the transmon qubit.}
\end{figure}

\section{Methods}

The experiments are conducted on a circuit quantum electrodynamics platform. The system consists of a three-dimensional (3D) cavity storage resonator, a planar stripline readout resonator, and a superconducting transmon qubit (in the following text qubit)~\cite{Koch2007}. A 3D cavity in a coaxial $\lambda/4$ stub geometry~\cite{reagor2016} is machined into a block of high-purity aluminum. The cavity is first chemically etched to remove surface damage from the machining and then subsequently annealed~\cite{kudra2020high}. The field distribution of the RF mode of the cavity is presented in Kudra et al.~\cite{kudra2020high}. A single-junction transmon qubit and the micro-strip readout resonator are fabricated on a silicon chip, which is then inserted into the 3D cavity [\figpanels{fig:setup}{c}{d}]. It is measured inside a dilution refrigerator with a base temperature below \unit[10]{mK}. 

The device is controlled with two input lines: one for driving the cavity and one for driving the readout resonator and the qubit. The measurement and control pulses are generated and read out with the microwave transceiver platform Vivace~\cite{Vivace_link}. Pulsed waveforms are sampled at a \unit[2]{GSa/s} rate and up-converted to microwave frequencies by mixing them with continuous-wave carrier tones. The signals are sent to the device through a series of attenuators and filters. The readout signal is amplified by a chain having a Josephson travelling-wave amplifier~\cite{macklin2015near} as the first amplifier, down-converted to an intermediate frequency, and acquired in the time domain at a sampling rate of \unit[2]{GSa/s} by Vivace.

The relevant part of the qubit-storage cavity system is described by the Hamiltonian~\cite{heeres2015cavity}
\begin{align}
\label{eq:hamil}
    H = \omega_c a^\dagger a - \dfrac{K_c}{2}(a^\dagger)^2 a^2 + \omega_{q} b^\dagger b \nonumber \\
    - \chi a^\dagger a b^\dagger b - \dfrac{\chi'}{2}(a^\dagger)^2 a^2 b^\dagger b .
\end{align}
Here $\omega_c$ and $\omega_q$ are the resonance frequencies of the cavity and the transmon qubit, respectively, $K_c$ is the Kerr nonlinearity of the cavity, $\chi$ is the dispersive shift between the cavity and the transmon qubit, $\chi'$ is a photon-number-dependent correction to the dispersive shift, $a^\dagger$ ($a$) is the creation (annihilation) operator of the cavity field, and, similarly, $b^\dagger$ ($b$) is the raising (lowering) operator of the qubit. When doing optimization or simulation we only consider the lowest two energy levels of the qubit. The Hamiltonian parameters and the coherence properties of the system are listed in~\appref{app:A}.

To prepare an arbitrary state in the cavity, one can apply an interleaved sequence of $n+1$ displacement operations and $n$ SNAP gates [\figref{fig:setup}(a)]. To implement this sequence, we introduce a two-step optimization routine [\figref{fig:setup}(b)]. 
The first optimization step takes the target state as an input and computes the amplitudes of the displacement gates ($\alpha_1$,$\alpha_2$,...,$\alpha_{n+1}$) and phases of the SNAP gates ($\vec{\theta}_1$,$\vec{\theta}_2$,...,$\vec{\theta}_n$), for a fixed number of $n$ SNAP gates (see~\appref{app:Optimization} for more details). The $i$th SNAP gate can be mathematically described as~\cite{heeres2015cavity,krastanov2015universal}:
\begin{align}
\label{eq:SNAP}
    S(\vec{\theta}_i) = \prod_{j=0}^{m}e^{{i\theta_{i,j}\ket{j}\bra{j}}}
\end{align}
and amounts to simultaneously applying the phase $\theta_{i,j}$ to the $j$th Fock state, with $j=0,1,\ldots,m$. In this work we have used $n=2,3$ and $m$ between 5 and 17, depending on which state is being prepared (see~\tabref{tab:table1}). For every state we prepared we try this step of the optimization for all $n$ smaller than the one chosen in this work. The coherent errors for smaller $n$ reduce the fidelity of the states. The incoherent errors further degrade the fidelity of the states when increasing the $n$ beyond the $n$ chosen in this work.
In the second step, we calculate the optimal pulse envelopes to implement the required SNAP gates $S(\vec{\theta}_i)$, based on the measured parameters of the Hamiltonian in \eqref{eq:hamil}, using the optimization tool Boulder Opal~\cite{qctrlsnap}. The tool solves for the unitary dynamics of the system in the presence of a driving term of the form $H_d=\gamma(t)b + \mathrm{h.c.}$, where $\gamma(t) = I(t)+iQ(t)$ is the complex control amplitude to be optimized. In Appendix~\ref{app:Scaling of the method}, we show how the method scales by simulating the fidelity of the Fock states $\ket{1}-\ket{10}$ created by SNAPs and displacements.

SNAP gates have been previously implemented by applying a superposition of $m$ pulses centred at frequencies $\omega_q + j \chi$, with the $j$th pulse rotating the qubit conditioned on the cavity being in Fock state $j$~\cite{heeres2015cavity}. However, using that approach, the minimum duration of the SNAP gate is many times longer than $1/\chi$, limited by the requirement that each pulse should condition the dynamics on a single Fock state. In our approach, the optimal control pulses reduce the duration of the SNAP gates by about 8 times, from \unit[4]{$\mu$s} down to \unit[500]{ns}. At the same time, optimal control pulses correct for the phases acquired during the gate duration that result from the small Kerr nonlinearity of the cavity mode, $K_c$. Details of the calibration of the Hamiltonian parameters and the cavity and qubit Rabi rates are given in \appref{app:A}. We realize displacement operations, $D(\alpha)$, by resonantly driving the cavity with pulses with a sine-squared envelope, and a total duration of \unit[50]{ns}, and a calibrated amplitude proportional to $\alpha$.
\begin{figure*}[t!]
\includegraphics[width=180mm]{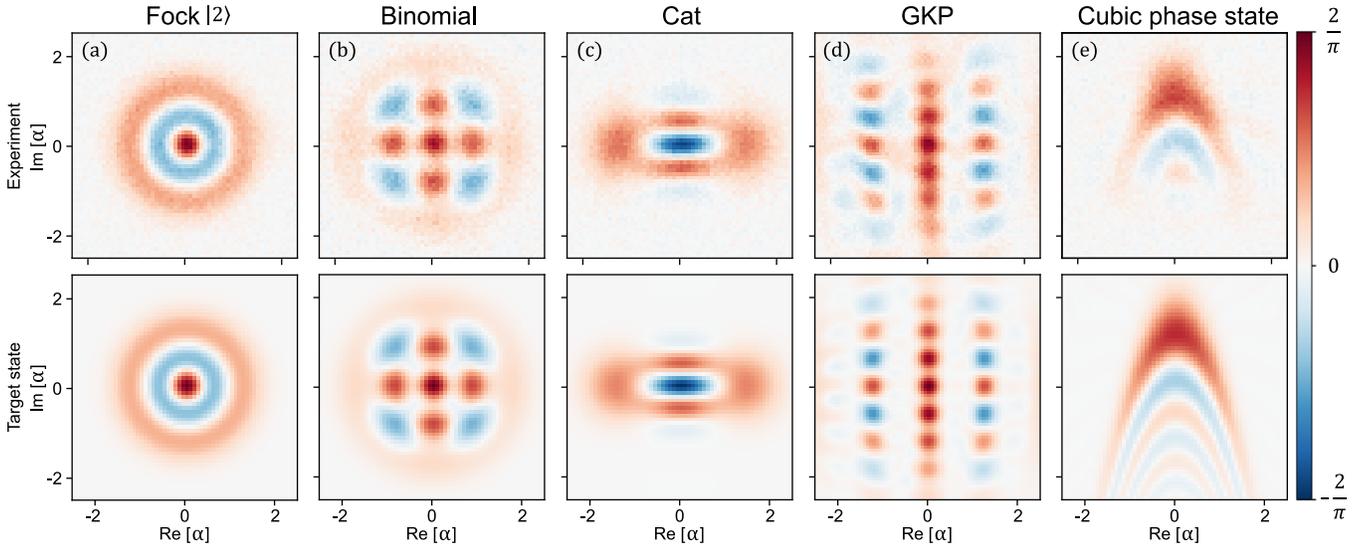}
\caption{\label{fig:wigners}Wigner functions of the states created by optimized SNAP displacement gates. Top row: experimental data. Bottom row: target states. From left to right: (a) Two-photon Fock state $|2\rangle$, (b) binomial state $(|0\rangle + |4\rangle)/\sqrt{2}$, and (c) an odd cat state $(|\alpha\rangle - |-\alpha\rangle)/\lambda$, with $\alpha = \sqrt{2}$ and $\lambda$ the normalization constant. These three states have 2 photons on average and were created with two SNAPs and three displacements. Next, (d) the GKP state with four photons on average and (e) the cubic phase state. The GKP state was prepared with three SNAPs and four displacements and the cubic phase state was prepared with three SNAPs and three displacements. Corresponding fidelities are listed in \tabref{tab:table1}.}
\end{figure*}

\begin{figure*}[t!]
\includegraphics[width=180mm]{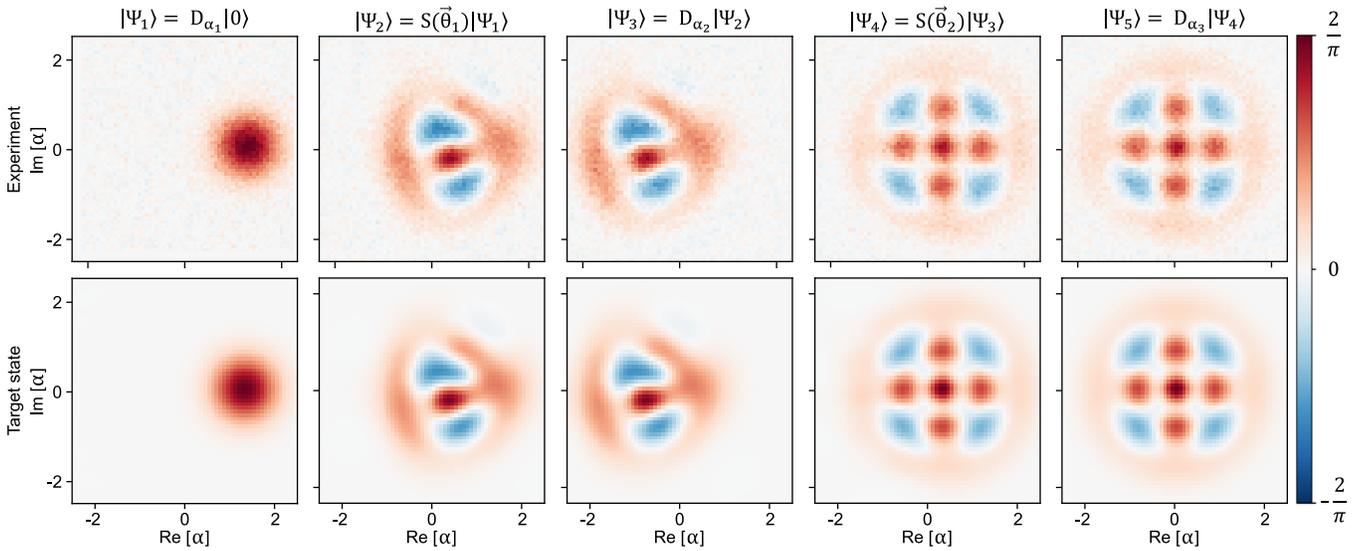}
\caption{\label{fig:wigners_binomial_steps}Wigner function after each gate in the sequence to create a binomial state $(|0\rangle + |4\rangle)/\sqrt{2}$. Top: experimental data. Bottom: target states. The fidelities after each gate are listed in \tabref{tab:table1}. The parameters of each SNAP and displacement gate are given in \tabref{tab:dispSNAPpar}.}
\end{figure*}
Following the state preparation sequence, we perform direct Wigner tomography of the prepared state~\cite{heeres2015cavity}. To do so, we apply a displacement of varying complex amplitude $\alpha$ followed by a Ramsey measurement at fixed time delay $1/2\chi$, which returns the photon parity of the cavity mode as the outcome of the qubit measurement. The Wigner function is then obtained as:
\begin{equation}
    W(\alpha) = \tfrac{2}{\pi}\mathrm{Tr}[D^\dagger(\alpha) \rho D(\alpha) \Pi],
\end{equation}
with $\Pi$ being the parity operator.
We reconstruct the density matrix of the target state from the Wigner function using a recently developed neural-network-based approach~\cite{Ahmed2021a, Ahmed2021b} which reconstructs the most likely density matrix of the state based on the measured Wigner tomography data.

\section{Results}

To showcase our ability to prepare arbitrary states of the cavity using displacement-SNAP gates, we prepare Wigner-negative states belonging to different classes of quantum states. We prepare the Fock state $|2\rangle$, the binomial state $(|0\rangle + |4\rangle)/\sqrt{2}$, and the cat state $(|\alpha\rangle-|-\alpha\rangle)/N$, where $\alpha = \sqrt{2}$ and $N$ is a normalization constant [\figpanels{fig:wigners}{a}{c}]. All three states have an average photon number of two and are prepared using two SNAP gates and three displacements. 
We also create a Gottesman-Kitaev-Preskill (GKP) state by applying three SNAPs and four displacements [\figpanel{fig:wigners}{d}] and the cubic phase state by applying three SNAPs and three displacements [\figpanel{fig:wigners}{e}] . The precise definitions of the targeted GKP and cubic phase states are given in Appendix~\ref{app:GKP_and_cubic}.

The fidelities of the prepared states are listed in~\tabref{tab:table1}
along with the number of applied displacement and SNAP gates, and compared to theoretical fidelities. The latter are obtained by numerically solving a master equation (see~\appref{app:simulations}) which takes into account the measured Hamiltonian parameters [\eqref{eq:hamil}], the applied SNAP and displacement gates, and the average measured decoherence and dephasing of both the qubit and the cavity.
The fidelities of the states with two photons on average (Fock, binomial, and cat) exceed 0.99 in the ideal case and they are limited by the coherence parameters of our system (to about 0.97). The fidelity of the Fock state $|2\rangle$ exceeds the predicted fidelity by 0.02, while the experimentally obtained binomial state has fidelity 0.03 lower than theoretical prediction. We can  ascribe about $\pm 0.01$ of discrepancies to fluctuations in the coherence time of the qubit~\cite{burnett2019decoherence}. Further, we estimate the reconstruction uncertainty to be 0.01 by bootstrapping. 
\begin{table}[t]
\caption{\label{tab:table1}
State fidelities for the generated states shown in Figs.~\ref{fig:wigners} and \ref{fig:wigners_binomial_steps}. $N_D$ ($N_S$) is the number of displacement (SNAP) gates we use. The variable $m$ is the highest Fock state that the SNAP gate applies a non-zero phase to. $\ket{\Psi_i}$ are the states captured after each gate in a sequence of preparing the binomial state. }
\begin{ruledtabular}
\renewcommand{\arraystretch}{1.25}
\renewcommand{\tabcolsep}{0.1cm}
\begin{tabular}{ccccc}
State & ($N_D$,$N_S$)& $m$ & Experiment & Theory\\
\hline   
Fock $\ket{2}$ & (3,2) & 6 & 0.99 & 0.97  \\ 
Binomial & (3,2) & 7 & 0.94 & 0.97   \\ 
Cat & (3,2) & 10& 0.96 & 0.97 \\ 
GKP & (4,3)& 17& 0.94 &  0.93 \\ 
Cubic phase & (3,3) & 10 & 0.87 & 0.92 \\ 
\hline
$\ket{\Psi_1} = D_{\alpha_1} \ket{0}$  & (1,0)& - & 0.99 & 0.99\\ 
$\ket{\Psi_2} = S(\vec{\theta}_1) \ket{\Psi_1}$  & (1,1) & 5& 0.96 & 0.98 \\ 
$\ket{\Psi_3} = D_{\alpha_2} \ket{\Psi_2}$  &(2,1)& -& 0.95 & 0.98 \\ 
$\ket{\Psi_4} = S(\vec{\theta}_2) \ket{\Psi_3}$ & (2,2)& 7 & 0.92 & 0.97 \\ 
\end{tabular}
\end{ruledtabular}
\end{table}
The GKP state, which has 4 photons on average, was created by three SNAP gates that add phases to all Fock states up to Fock number 17. The corresponding displacement-SNAP sequence would create a GKP state with fidelity 0.99 in the ideal case, but with loss in our system that number drops to 0.93. This is in good agreement with the experimentally obtained 0.94.
The cubic phase state created by three SNAPs and three displacements has fidelity 0.97 to the targeted cubic phase state with cubicity $\gamma = -0.106$ and a squeezing parameter
$\zeta = 0.5$ (see Appx.~\ref{app:GKP_and_cubic} and \figref{fig:wigners}) in the ideal case. After taking loss into account, the simulated fidelity is 0.92. The achieved experimental fidelity is 0.87, when taking first 10 Fock states into account. We justify taking this cutoff by the fact that more than 0.999 of the Fock state population of state created by the SNAPs and displacements is within the first 10 Fock states. The Wigner functions of the reconstructed state, and further discussion is given in Appendix~\ref{app:reconstruction_cubic}.

An advantage of the SNAP-displacement approach to state preparation is that the contribution of each individual step of the sequence to the final fidelity can be independently analyzed by taking a Wigner tomography of intermediate states after each step. This would potentially allow us to identify weak links in the sequence and optimize the gates individually. Such a series of ``Wigner snapshots'' taken after each step of the sequence to prepare the binomial state $(|0\rangle + |4\rangle)/\sqrt{2}$ is presented in~\figref{fig:wigners_binomial_steps}. The fidelities of intermediate states monotonically decrease as the sequence progresses~(see \tabref{tab:table1}). SNAP gates, being 10 times longer than the displacement gates, are responsible for most of the infidelity. 
\begin{figure}[t]
\includegraphics[width=90mm]{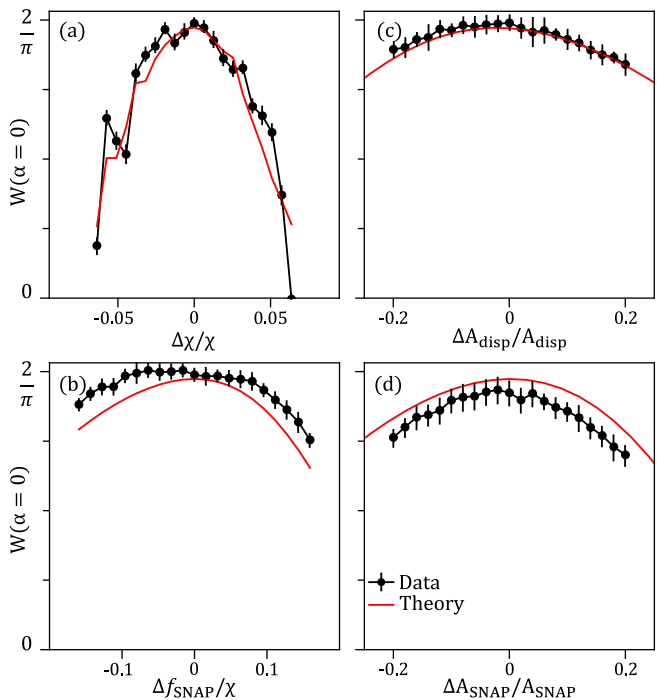}
\caption{\label{fig:sensitivity}
Sensitivity of the displacement-SNAP sequence to different calibration parameters.
Wigner function at the origin for a two-photon Fock state vs (a) dispersive shift, $\chi$, (b) modulation frequency of SNAP gates $f_{\rm SNAP}$, (c) amplitude of the displacement pulses, $A_{\rm disp}$, and (d) amplitude of the SNAP pulses, $A_{\rm SNAP}$. Dots: experiment. Solid line: theory. The error bars represent one standard deviation of 20 repetitions of each experimental point.}
\end{figure}

\begin{figure}[t]
\includegraphics[width=90mm]{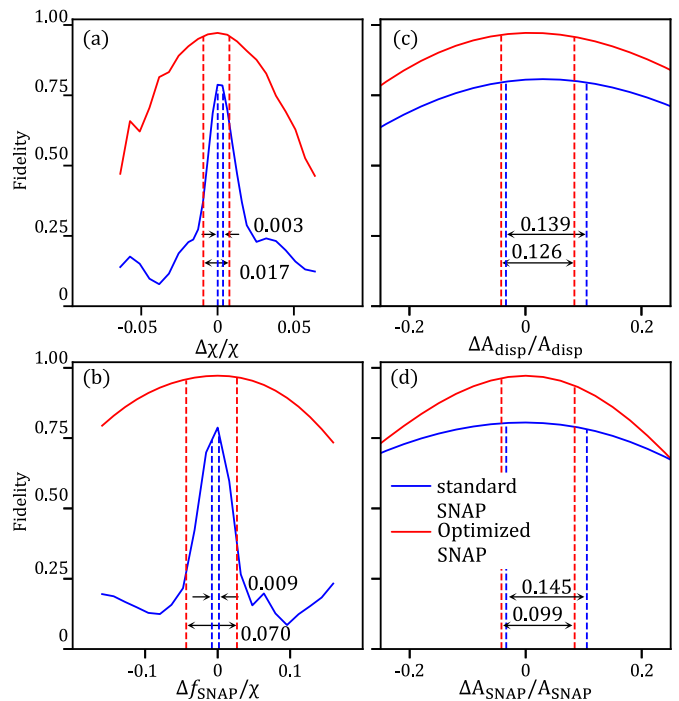}
\caption{\label{fig:theory_fidelities} Simulated fidelities for optimized and ``standard'' SNAP gates vs the same calibration parameters as in \figref{fig:sensitivity}. The distance between the dashed lines represents the parameter span where the fidelity is within \unit[1]{\%} of the maximum. Optimized SNAP gates are identical to those in \figref{fig:sensitivity}, while ``standard'' SNAP gates are a superposition of \unit[4]{$\mu$s} long pulses centered at photon-number-dependent qubit frequencies~\cite{heeres2015cavity}.}
\end{figure}
We also probe the sensitivity of the state preparation sequence to different calibration parameters that are fed to our pulse optimizer (\figref{fig:sensitivity}). To do so, we prepare the two-photon Fock state with three displacements and two SNAPs [same as \figpanel{fig:wigners}{a}], and measure the Wigner function at the origin, $W(0)$, as a proxy for its preparation fidelity. This state has even parity; hence the value of the Wigner function at the origin has the maximum value the Wigner function can take ($2/\pi$). Thus a lower value of the Wigner function must be associated with a decrease in fidelity, as confirmed by simulations. 
To probe the sensitivity to the dispersive shift $\chi$, we vary its value fed to the optimizer by $\pm0.07$ around its nominal value. For each value, we generate a new pulse sequence, execute it, and measure $W(0)$ [\figpanel{fig:sensitivity}{a}]. In addition, we explore variations in the carrier frequency of the SNAP pulses $f_{\rm SNAP}$ [\figpanel{fig:sensitivity}{b}], the amplitude of the displacement pulses $A_{\rm disp}$ [\figpanel{fig:sensitivity}{c}], and the amplitude of the SNAP pulses $A_{\rm SNAP}$ [\figpanel{fig:sensitivity}{d}]. 
The theory lines plotted in red in \figref{fig:sensitivity} are obtained by simulating an open system given by the qubit and the cavity with the measured average decay and dephasing rates. The only fitting parameter is the offset in the x-axis. The offsets tell us how far from the optimal calibration point our applied pulses are. The displacement amplitude offset is -0.03 and the rest are less than 0.01. The theory lines capture well the parabolic decrease in $W(0)$ as the pulse parameters deviate from the optimum; however, they  deviate from the experimental data by a constant factor in \figpanel{fig:sensitivity}{b} and (d). We attribute these deviations to fluctuations in the coherence parameters.
%

Further, we compare the sensitivity of ``standard'' and optimized SNAP gates. To do this, we simulate how the state preparation fidelity of Fock state $|2\rangle$ depends on the parameters $\chi$, $A_{disp}$, $f_{SNAP}$ and $A_{SNAP}$ (like in \figref{fig:sensitivity}), for both our optimized and standard SNAP gates (\figref{fig:theory_fidelities}). We define a standard SNAP gate $S(\theta_0,\theta_1,...,\theta_m)$ as a sum of $m$ pulses at frequencies $\omega_q-i\chi, i\in\{0,1,...,m\}$, where $m$ is the maximum Fock level the SNAP gate adds phase to. At each of the frequencies, two consecutive $\pi$ pulses (2~$\mu$s long) with axes of rotation $\theta_i$ apart are sent~\cite{heeres2015cavity}. The distance between the dotted lines represents the parameter range which is within \unit[1]{\%} of the maximum fidelity.
The fidelity of the state made by the standard SNAPs is 5 times more sensitive to calibration of dispersive shift $\chi$ [\figpanel{fig:theory_fidelities}{a}] and 7 times more sensitive to fluctuations in qubit frequency [\figpanel{fig:theory_fidelities}{b}]. This robustness to fluctuations in qubit frequency and dispersive shift is a feature of the optimized SNAP pulses. The displacement amplitude can be about $\pm 0.06$ from the optimal value for both the optimized and standard SNAP pulses [\figpanel{fig:theory_fidelities}{c}]. The displacement pulses are identical for both state preparation cases, so we would expect them to have similar sensitivity to amplitude of the displacement pulses. Finally, the amplitude of the SNAP pulses can vary $\pm 0.05$ for optimized SNAP pulses and $\pm 0.07$ for standard SNAP pulses [\figpanel{fig:theory_fidelities}{d}]. An important thing to note here is that the fidelity close to the maximum has a parabolic dependence on all of the calibration parameters. This allows us to independently further fine-tune these parameters.

\section{Discussion}

We have shown that only a few optimized SNAP and displacement gates provide a well-controlled method for generating a range of complex, highly nonclassical states relevant for quantum computation. We tested the performance of this state preparation scheme by generating Fock, binomial, cat, GKP, and cubic phase states, achieving the first experimental implementation of the latter. While binomial, cat and GKP states find their natural application in fault-tolerant quantum computation~\cite{PhysRevX.10.011058, Grimsmo2021}, the cubic phase state increases the fidelity in continuous-variable teleportation~\cite{Zinatullin2021}, can be used to implement a T-gate on encoded GKP states~\cite{Gottesman2001, Konno2021}, and promotes the set of Gaussian operations to arbitrary Hamiltonian evolution yielding CV universality~\cite{Lloyd1999, Gottesman2001} and enabling the study of  CV-NISQs algorithms~\cite{hillmann2020}. 

Further, we showed that the optimized SNAP gates have parabolic dependence in fidelity with respect to perturbations of relevant calibration parameters close to the maximum fidelity. Furthermore, this preparation scheme is robust to fluctuations in both displacement amplitude ($\Delta A_{\rm disp}/A_{\rm disp} = \pm 0.06$) and in the amplitude ($\Delta A_{\rm SNAP}/A_{\rm SNAP} = \pm 0.05$) and frequency ($\Delta f_{\rm SNAP}/\Delta \chi = \pm 0.04$) of the SNAP drive. Compared to the SNAP gates in Ref.~\cite{heeres2015cavity}, the optimized SNAP gates are about 8 times shorter, 5 times less sensitive to the calibration of dispersive shift $\chi$, and 7 times less sensitive to fluctuations of qubit frequency. To shed light on the origin of this robustness, it would be interesting to compare the optimizer we used (Boulder Opal from QCTRL~\cite{qctrlsnap}) against GRAPE~\cite{heeres2017implementing} to create both optimized SNAPs and optimal control pulses in order to see how robust they are to variations in Hamiltonian parameters. Another possibility is to explore whether the optimized SNAP gates could be made path independent by using the third level of the transmon qubit~\cite{reinhold2020error,Path_independant} and posing additional constraints on the optimizer. 

The methods presented here are also suited for initializing the states of interacting harmonic modes, as well as potentially applying logical gates on the logical states~\cite{fosel2020efficient}. The 8 times shorter SNAP gate time brings the implementation of the quantum simulations with bosonic systems closer to realization~\cite{kurkcuoglu2021quantum}. Improving the coherence time of the qubit is one obvious way to improve the fidelity of all of the created states. Our results are relevant for any implementation of continuous-variable quantum computing with qubit-oscillator systems in the strong dispersive regime, including 3D microwave cavities~\cite{reagor2016,chakram2020multimode}, superconducting resonators~\cite{lescanne2020}, and acoustic resonators coupled to superconducting qubits~\cite{satzinger2018quantum,kervinen2020sideband,wollack2021,von2021parity}.

\begin{acknowledgments}

We would like to thank Mats Myremark and Lars J\"onsson for machining the cavity, Gustav Gr\"annsj\"o and Johan Blomberg for their help with the Vivace driver. We would further like to thank Andre Carvalho and Viktor Perunicic from Q-CTRL for their help with the optimization and simulation software Boulder Opal. The simulations and visualization of the quantum states were performed using QuTiP~\cite{Johansson2012,Johansson2013}, NumPy~\cite{Harris2020array}, and Matplotlib~\cite{Hunter2007}. The automatic differentiation tools TensorFlow~\cite{TensorFlow2015} and Jax~\cite{Jax2018} were used in state reconstruction and optimization. This work was supported by the Knut and Alice Wallenberg foundation via the Wallenberg Centre for Quantum Technology (WACQT) and by the Swedish Research Council. M. Kervinen acknowledges support from the Emil Aaltonen Foundation. The chips were fabricated in Chalmers Myfab cleanroom. We acknowledge IARPA and Lincoln Labs for providing the TWPA used in this experiment.

\end{acknowledgments}

\appendix

\setcounter{table}{0}
\renewcommand{\thetable}{A\arabic{table}}
\setcounter{figure}{0}
\renewcommand{\thefigure}{A\arabic{figure}}

\section{Hamiltonian parameters}
\label{app:A}


\begin{table}[h]
\caption{\label{tab:Hamilton_param}
Parameter values for the Hamiltonian in \eqref{eq:hamil} and measured coherence times. Qubit relaxation and decoherence times vary over time and are given with two sigma variance. }
\begin{ruledtabular}
\begin{tabular}{ccc}
Parameter & Symbol & Value \\ 
\hline
Qubit frequency & $\omega_q/2\pi$ & \unit[6.2674]{GHz} \\
Cavity frequency & $\omega_c/2\pi$ & \unit[4.428018]{GHz} \\ 
Resonator frequency & $\omega_r/2\pi$ & \unit[7.2859]{GHz} \\
Qubit-cavity disp.~shift & $\chi_{qc}/2\pi$& \unit[3.14]{MHz}\\
Qubit-resonator disp.~shift & $\chi_{qr}/2\pi$& \unit[1.7]{MHz} \\
Cavity Kerr coeff. & $K_c/2\pi$& \unit[6]{kHz}  \\
Qubit anharmonicity & $\alpha_{q}/2\pi$& \unit[-300]{MHz} \\
Qubit-cavity disp.~shift corr. & $\chi'_{qc}/2\pi$& \unit[25]{kHz}\\
Qubit relaxation time & $T_{1q}$ & \unit[35 $\pm$ 14]{$\mu$s} \\
Qubit decoherence time & $T_{2q}$ & \unit[27 $\pm$ 8]{$\mu$s} \\
Cavity relaxation time & $T_{1c}$ & \unit[248]{$\mu$s} \\
\end{tabular}
\end{ruledtabular}
\end{table}

\begin{table}[h]
\caption{\label{tab:dispSNAPpar}
Gate parameters used to create the states in \figpanels{fig:wigners}{a}{c}.}
\begin{ruledtabular}
\begin{tabular}{ccc}
State~~ & \makecell{Gate\\parameters} & Values\\
\hline
\multirow{3}{*}{\makecell{Fock\\state\\$|2\rangle$}}& $\alpha_1$, $\alpha_2$, $\alpha_3$& 1.390, -0.494, 0.622\\
&$\vec{\theta}_1$&\makecell{(2.049, -0.654, 1.130, -1.106)}\\
&$\vec{\theta}_2$&\makecell{(0.003, 1.592, 0, -0.869, 0, -0.234,\\ 0.067)}\\
\hline
\multirow{3}{*}{\makecell{Binomial\\state}}& $\alpha_1$, $\alpha_2$, $\alpha_3$& 1.304, -1.144, -0.291\\
&$\vec{\theta}_1$&\makecell{(0.111, -1.535, 0.377, 0, -0.836,\\ -0.856)}\\
&$\vec{\theta}_2$&\makecell{(0.404, 0, 0, -0.777, -0.976, 2.083,\\ -1.087, 1.552)}\\
\hline
\multirow{3}{*}{\makecell{Cat\\state}}& $\alpha_1$, $\alpha_2$, $\alpha_3$&1.373, -0.614, 0.529\\
&$\vec{\theta}_1$&\makecell{(1.277, -0.302, -1.906, 0.093, -1.161,\\ 0.463, -0.569, 0.286)}\\
&$\vec{\theta}_2$&\makecell{(0.650, 2.109, 0, 0.770, 0.392, 0.272, 0,\\ -0.132, -0.248, -0.251, -0.027)}\\
\end{tabular}
\end{ruledtabular}
\end{table}

The system parameters are listed in \tabref{tab:Hamilton_param}. The different parameters such as the dispersive shift $\chi$, the correction to the dispersive shift $\chi'$, and the Kerr nonlinearity of the cavity are measured following the calibration experiments in Ref.~\cite{Reinhold_PhD}. Besides the Hamiltonian parameters from \eqref{eq:hamil}, the pulse optimization is constrained by the maximum Rabi rate, the sampling frequency, and the cut-off for the digital low-pass filter function. The low-pass cut-off is \unit[40-80]{MHz} depending on the target state which ensures the drive can be implemented on our hardware without exciting unwanted transitions. Since the anharmonicity of the qubit $\alpha_q$ is much larger than the filter cut-off, we need to take only two qubit levels into account in \eqref{eq:hamil}. The conversion between the SNAP pulse amplitudes in frequency units and the control voltages output by the microwave platform Vivace is calibrated through standard Rabi frequency measurement. The maximum Rabi rate is ensured to be under \unit[30]{MHz} to be below the maximum output amplitude of our instrument and to avoid any non-linear effects of strong driving. Finally, the sampling frequency was chosen to be \unit[1]{GSa/s} (i.e., 500 points for a \unit[500]{ns} pulse).

\section{Wigner-function reconstruction}
\label{app:reconstruction}
In order to obtain the density matrix and to estimate fidelity of the target state, we reconstruct the density matrix from the Wigner tomography measurement using a recently developed neural-network-based approach~\cite{Ahmed2021a, Ahmed2021b}, which  reconstructs  the density matrix that most faithfully reproduces the Wigner tomography of the state. The fidelity is then given by 
\begin{equation}
F(\rho,\sigma) = \left(\mathrm{Tr}\sqrt{\sqrt{\rho}\sigma \sqrt{\rho}}\right)^2,    
\end{equation}
where $\rho$ and $\sigma$ are the density matrices for the target and the reconstruction. For the reconstructed density matrix, we use a Hilbert space size of 32. 

We measure the Wigner function by using an unconditional $\pi/2$-pulse followed by a waiting time $1/(2\chi)$ and finally either $\pi/2$ or a $-\pi/2$ pulse for every Wigner point $W(\alpha)$. We subtract the two readout amplitudes to get the Wigner amplitude $A(\alpha)$. In addition, we measure the readout amplitudes for the same sequences without the waiting time $1/(2\chi)$ for the center of the Wigner function to get an amplitude $B$. The scaled Wigner function we report is $(\alpha)/B$. This whole procedure is similar to the one detailed in Ref~\cite{chakram2020multimode}. It is designed to compensate for the loss of contrast in the Wigner function when measuring at large Fock states. However, in our experimental device we had a small readout linewidth of $\approx 50 kHz$ which further amplified the sensitivity to amplitude variations of the readout signal. This might explain the remaining discrepancies in the fidelities between the theoretically and the experimentally generated states.


\section{Master-equation simulations}
\label{app:simulations}
To obtain the theoretical fidelities of our target states, we simulate the system dynamics using a master equation with the known decoherence sources, and the applied pulse waveforms. The Lindblad master equation is defined as 
\begin{align}
\label{eq:me}
    \dfrac{\partial}{\partial t} \rho(t) &= -i[H(t),\rho(t)] + \nonumber \\
    &\left( \dfrac{1}{T_{1c}}\mathcal{D}[a]+\dfrac{1}{T_{1q}}\mathcal{D}[b]+\dfrac{1}{T_{\phi}}\mathcal{D}[b^\dagger b] \right)\rho(t),
\end{align}
where $T_{1q}$ and $T_{\phi}$ are the decay and the dephasing of the qubit, respectively, whereas $T_{1c}$ is the decay of the cavity mode. $\mathcal{D}$ is the Lindblad superoperator:
\begin{equation}
\mathcal{D}[a]\rho = a \rho a^\dagger -\dfrac{1}{2} \{ a^\dagger a, \rho \},    
\end{equation}
which describes the decoherent procesess in the system evolution.

\section{Optimization using gradient descent}
\label{app:Optimization}

The parameters of the displacement and SNAP gates that are used to create the states in the main text are given in \tabref{tab:dispSNAPpar}. We obtain these parameters by using gradient-based optimization of the cost function:
\begin{eqnarray}
\label{eq:cost}
    \mathcal L(\alpha_i, ..., \vec \theta_j, ...) &=& [1 - F(\Psi_{\text{target}}, \Psi(\alpha_i, ..., \vec \theta_j, ...))] \nonumber \\
    &+& \lambda \sum_{i, m}|\theta_{m}^{i}| ,
\end{eqnarray}
where $\Psi(\alpha_i, ..., \vec \theta_j, ...)$ represents a variational state. This variational state is prepared by applying displacements and SNAP gates to the vacuum state. We use the fidelity $F(\Psi_1, \Psi_2)$ between the two states to construct our loss function. The LASSO regularization~\cite{Tibshirani1996} term $\lambda \sum_{i, m}|\theta_{m}^{i}|$ is used to keep the SNAP parameters sparse.

\begin{figure*}[t!]
\includegraphics[width=180mm]{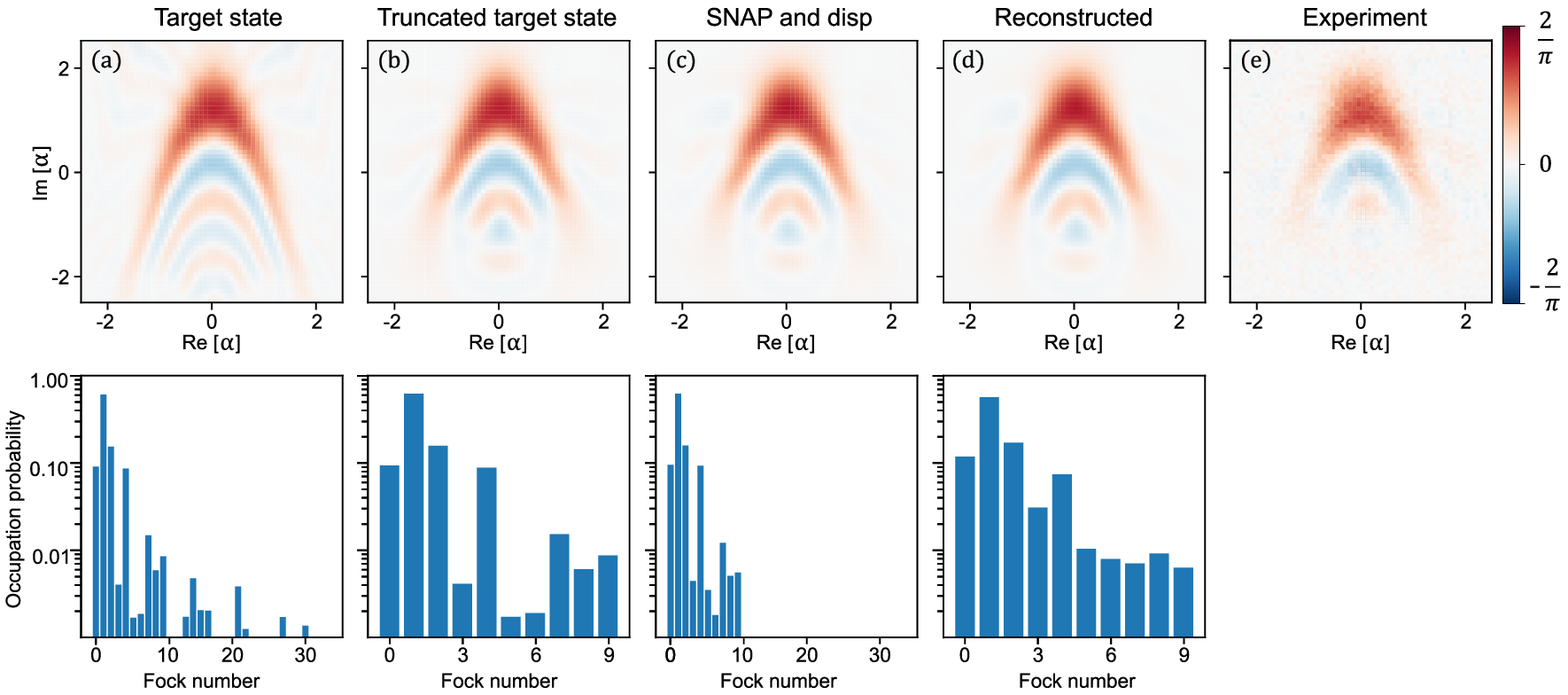}
\caption{\label{fig:cubic_review} Comparison of the Wigner functions and occupation probability of, from left to right, the target cubic phase state, the target cubic phase state truncated to the first 10 Fock states, the state prepared by three SNAPs and three displacements, the reconstructed state limited to the first 10 Fock states, and the Wigner function of the experimental data.}
\end{figure*}

In Ref.~\cite{fosel2020efficient}, the authors demonstrated that only $3$ to $4$ SNAP gates were sufficient for relevant experimental implementation --- an improvement from the $\sim 50$ gates in previous approaches. In our work, we fix the number of SNAP-displacement operations and then use gradient descent to minimize~\eqref{eq:cost} starting from a random set of displacement and SNAP parameters. Our approach differs from that of Ref.~\cite{fosel2020efficient} in that we directly optimize the displacement and SNAP parameters using gradient-descent by assuming a fixed length for the gate sequence. However the method in Ref.~\cite{fosel2020efficient} uses a two step process where in the first step, a heuristic is used to determine a reasonable length for the gate sequence and find an approximate initialization for the parameters. In the next step, the parameters are fine tuned with gradient-based optimization.

Since in our experiment, we can only apply a limited number of gates, we fixed the length of the sequence directly and only optimize the parameters akin to the fine-tuning step of Ref.~\cite{fosel2020efficient}. We observed that using appropriate hyperparameters and regularization of the loss function, we can directly obtain the gate parameters starting from a random initialization.

\section{GKP and cubic phase state}
\label{app:GKP_and_cubic}
We consider the finite GKP state~\cite{Gottesman2001, Albert2018, campagne2020quantum, Ahmed2021b} characterized by the real parameter $\sigma \in [0, 1]$ and defined in the Fock basis as,
\begin{equation}
\ket{\psi^{{\sigma, \mu}}_{\texttt{GKP}}} = \sum_{\alpha \in \mathcal K(\mu)} e^{-\sigma^2 |\alpha|^2} e^{- i \rm{Re} [\alpha ] \rm{Im} [\alpha]} D(\alpha)\ket 0,
\end{equation}
where $D(\alpha)$ is the displacement operator with the complex amplitudes $\alpha$ are chosen in a grid $\mathcal K(\mu) = \sqrt{\frac{\pi}{2}} (2 n_1 + \mu)) + i \sqrt{\frac{\pi}{2}}n_2$, $\mu \in \{0, 1\}$ represents the logical $0$ or $1$ encoding, and $n_1, n_2$ are integers in $[-8, 8]$. We take the Fock space cutoff as $25$ with $\sigma = 0.35$ and $\mu = 0$.

The cubic phase state is defined as
\begin{equation}
\ket{\psi^{{\gamma, \zeta}}_{\texttt{cubic}}} = e^{i \gamma q^3} S(\zeta) \ket {0} ,
\end{equation}
where $q = (a + a^{\dagger})$ is defined using the ladder operators $a, a^{\dagger}$ and $S(\zeta)$ is the single-mode squeezing operator
\begin{equation}
S(\zeta) = e^{\frac{1}{2} \mleft(\zeta^* a^2 - \zeta {a^{\dagger}}^2 \mright)}.
\end{equation}
We use a cubicity $\gamma=-0.106$ and a squeezing parameter $\zeta = 0.5$ to generate our cubic phase state. An additional displacement of $\beta = 1.5i$ is applied to move the state in order to capture more features in the Wigner function plot.

\section{Reconstructing the cubic phase state}
\label{app:reconstruction_cubic}

The cubic phase state that we targeted [\figpanel{fig:cubic_review}{a}] has a total occupation probability of 0.97 in the first 10 Fock states. When truncated to the first 10 Fock states [\figpanel{fig:cubic_review}{b}], the Wigner function loses a few of the characteristic regions of alternating positive and negative values. The state created by applying three SNAPs and three displacements [\figpanel{fig:cubic_review}{c}] has more than 0.999 occupation probability within the first 10 Fock states. Its Wigner function closely resembles both the truncated target state and the reconstructed [\figpanel{fig:cubic_review}{d}] and experimental [\figpanel{fig:cubic_review}{e}] Wigner functions. Taking this into account, we justify allowing the reconstruction method to take only first 10 Fock states into account.

\section{Considerations on ``standard'' SNAP gates}
\label{app:Standard SNAP}

\begin{figure}[]
\includegraphics[width=90mm]{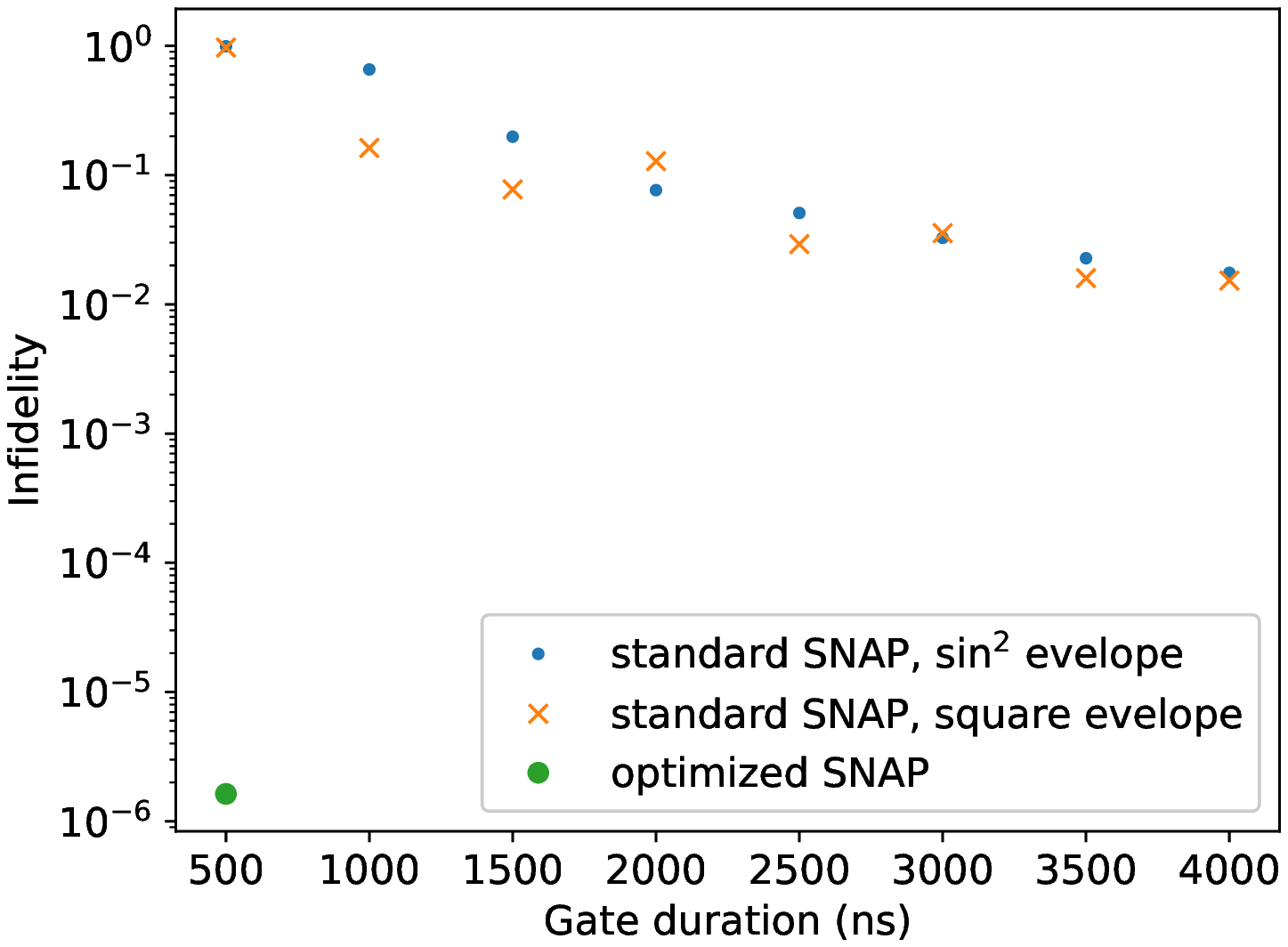}
\caption{\label{fig:length_of SNAP} Simulated comparison of the infidelity of the state $S(\vec{\theta_1})\ket{\alpha_1}$ created with standard or optimized SNAP gates of different lengths. $\alpha_1 = 1.39$ and $\theta_1 = (2.049, -0.654, 1.130, -1.106, 0, 0, ... )$ are the parameters of the first displacement and SNAP gates in the sequence making the Fock state $\ket{2}$. This simulation is performed without taking loss into account.}
\end{figure}

In order to find a fair length of the standard SNAP~\cite{krastanov2015universal,heeres2015cavity} to compare with the optimized SNAP, we perform a simulation that does not take loss into account and where we assume all the Hamiltonian parameters we measured are exact. We simulate the fidelity of the state $S(\vec{\theta_1})\ket{\alpha_1}$, where $\alpha_1 = 1.39$ and $\theta_1 = (2.049, -0.654, 1.130, -1.106, 0, 0, ... )$ are the parameters of the first displacement and SNAP gates in the sequence of making the Fock state $\ket{2}$. Results for both the $\sin^2$ and square envelopes of the standard SNAP, as well as the optimized SNAP that was used to create the Fock state $\ket{2}$ in the main text (Figs.~\ref{fig:wigners} and \ref{fig:sensitivity}), are summarized in \figref{fig:length_of SNAP}. The infidelity of the state created with the standard SNAP (no matter the envelope), which is \unit[4000]{ns} long, is 0.015. This is comparable to the decoherence error in our system that we have during the \unit[500]{ns} duration of the optimized SNAP. This is how we chose the \unit[4000]{ns} standard SNAP to compare with in the main text (\figref{fig:theory_fidelities}). The optimized SNAP induces four orders of magnitude smaller coherent error compared to the \unit[4000]{ns} standard SNAP provided that the Hamiltonian parameters we measure are exact.

\section{Scaling of the method (Simulating Fock states up to Fock state 10)}
\label{app:Scaling of the method}

\begin{figure}[]
\includegraphics[width=90mm]{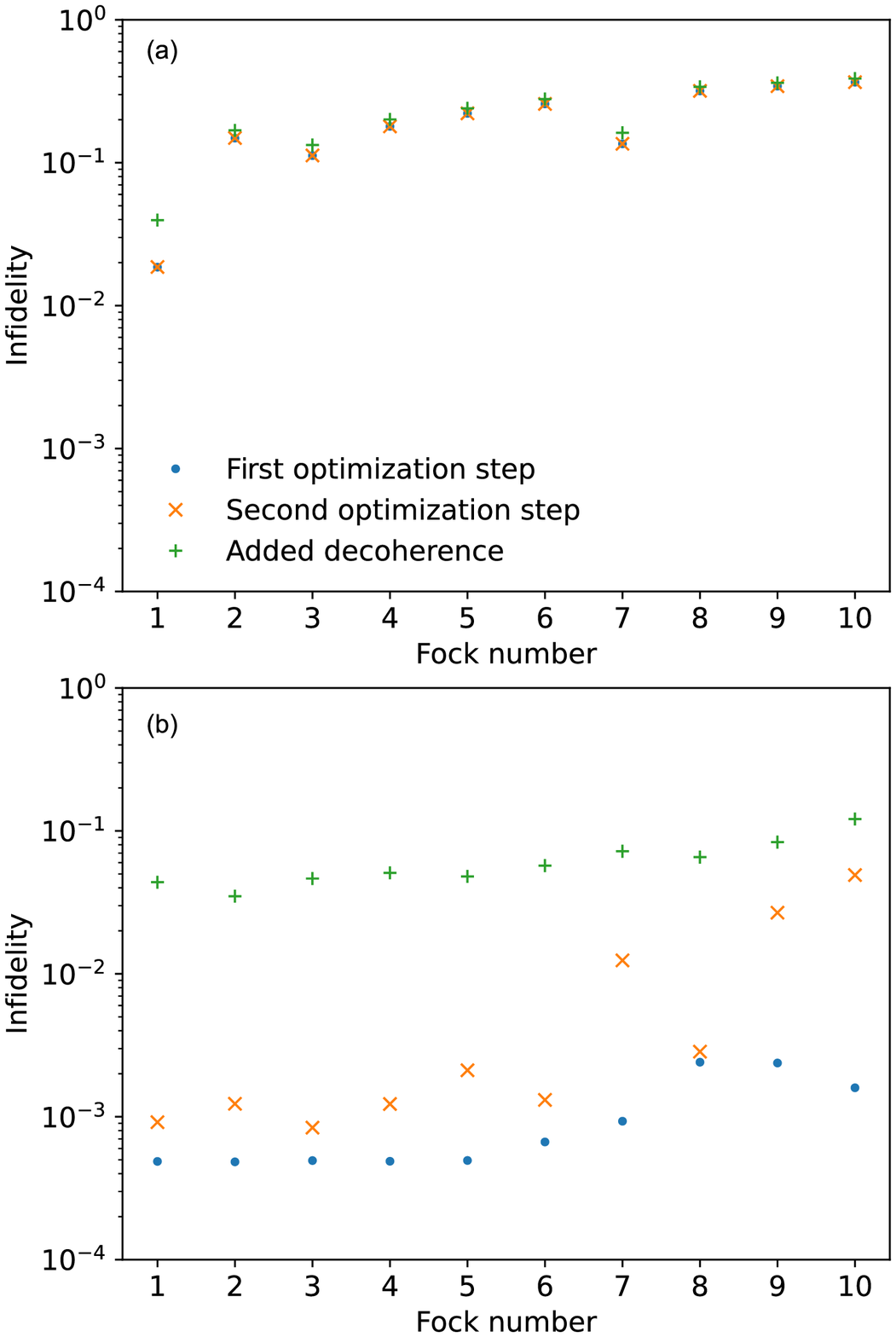}
\caption{\label{fig:fock dependance} Simulated infidelity of Fock states prepared using (a) one and (b) two SNAP gates. }
\end{figure}

We simulated preparation of Fock states from $\ket{1}$ to $\ket{10}$ using two displacements and one SNAP [\figpanel{fig:fock dependance}{a}] and using three displacements and two SNAPs [\figpanel{fig:fock dependance}{b}]. We simulated the fidelity after the first optimization step (finding the parameters of displacements and SNAPs), after the second optimization step (optimizing SNAP envelopes), and finally took loss into account.

Using only one SNAP gate, the fidelity of the Fock states goes from 0.98 for $\ket{1}$ down to 0.63 for $\ket{10}$ (after both the first and second optimization steps). The fidelity is limited by the first optimization step, where one SNAP gate is not enough to achieve better fidelities. Taking loss into account, the fidelities drop by 0.02. In order to create higher Fock states, the SNAP gate has to address more and higher Fock states. The maximum Fock number that the SNAP addresses to create $\ket{1}$ to $\ket{10}$ is [0, 10, 11, 10, 11, 12, 15, 15, 16, 17].

Using two SNAP gates, the fidelity of the Fock states $\ket{1}$ to $\ket{7}$ is more than 0.999 after the first optimization step (we limit the optimizer to stop if it reaches fidelity 0.999 for all states). Fock states $\ket{8}$ to $\ket{10}$ have fidelities of 0.998 after the first optimization step. The fidelity after the second optimization step in this case drops by less than 0.001 for $\ket{1}$; the drop increases to 0.04 for $\ket{10}$. The SNAP gates influence more Fock states (the maximum Fock number that the SNAP gates address to create $\ket{1}$ to $\ket{10}$ is [6, 8, 9, 13, 15, 16, 17, 20, 20, 21]). The second optimization step introduces more and more coherent errors in the SNAP gates that create higher Fock states. Taking loss into account, the fidelities drop by around 0.05. This means $\ket{1}$ has a fidelity of 0.95 and $\ket{10}$ has fidelity 0.88.

The simulation time for the first simulation step is longer for larger Fock states, but it is at most a few minutes on a standard desktop computer. The second optimization step is run in the Q-CTRL remote server, and the simulation time spans from a few minutes for low Fock states up to an hour for $\ket{10}$.

\bibliography{apssamp}

\end{document}